# The Statistical Physics of Cities

Marc Barthelemy[1,2]
[1] Institut de Physique Théorique (IPhT/CEA), Orme-des-Merisiers, 91191 Gif-sur-Yvette, France
[2] Centre d'Analyse et de Mathématique Sociales (CAMS/EHESS), 75006 Paris, France
Email: marc.barthelemy@ipht.fr

**Abstract:**
Challenges due to the rapid urbanization of the world — especially in emerging countries — range from an increasing dependence on energy, to air pollution, socio-spatial inequalities, environmental and sustainability issues. Modelling the structure and evolution of cities is therefore critical because policy makers need robust theories and new paradigms for mitigating these problems. Fortunately, the increased data available about urban systems opens the possibility of constructing a quantitative 'science of cities' with the aim of identifying and modelling essential phenomena. Statistical physics plays a major role in this effort by bringing tools and concepts able to bridge theory and empirical results. This Perspective illustrates this point by focusing on fundamental objects in cities: the distribution of urban population, segregation phenomena and spin like models, the polycentric transition of the activity organization, energy considerations about mobility and models inspired by gravity and radiation concepts, $CO_2$ emitted by transport, and finally scaling that describes how various socio-economical and infrastructures evolve when cities grow.

## I. Introduction

Cities are seen around the world and through history, suggesting 'universal' reasons for their existence. As concentrations in space, they reduce the distance between individuals, simplify the exchange of ideas, goods and services, and allow for specialization and sharing of skills and energy resources. However, cities have drawbacks: spatial localization leads to increased housing costs, traffic congestion, urban sprawl, pollution and other environmental problems. Devising and assessing mitigation strategies for these problems requires an understanding of cities and their evolution. This is especially true because cities in emerging countries are growing rapidly: it is estimated that by 2050 the population of urban areas worldwide will increase by 2.5 billion people, with 90% of this increase in Africa and Asia [1].

Most previous studies of cities come from the social sciences; geography [2] and economics [3] deal with quantitative aspects of cities. However, despite a large number of studies, it is fair to say that we do not yet have simple theoretical models for cities with predictions in agreement with empirical data. Fortunately, this situation is about to change thanks to two revolutions: complex systems and data.

Most researchers agree that a signature of complex systems is the emergence of collective non-trivial behaviours from elementary constituents — often many of them. Cities naturally fall in the category of complex systems: they have a large number of constituents (such as individuals, institutions, governments) that interact with each other, leading to observable large-scale structures that evolve in time and space. Structures that can be at least partially thought of as emerging collective effects include critical infrastructures, clustering of different types of areas (such as residential or business), social segregation, congestion and car traffic.

There remains a legitimate question [4, 5] of what a science of cities could be. This problem, recurrent in the study of complex systems, is of great importance at a time when data-driven approaches such as machine learning encounter many successes. As discussed for complex systems in general [6], a metric is needed for assessing the success of a theory of cities. Although I do not believe that a 'theory of everything' is possible at this point for urban systems, certain aspects can be quantitatively addressed. Quantities such as the amount of $CO_2$ emitted by cities, the amount of energy used or waste produced and the structure of mobility patterns are more specialized questions but a model that could explain observations would be of great value. Not everything lies in predicting numerical values — which is of course important — but also in identifying the relevant mechanisms and parameter combinations. For example, a celebrated empirical study [7] suggested that gasoline consumption is governed by population density, but a theoretical framework is needed to understand if this is the correct control parameter and to assess the importance of other parameters such as the public transit density or the area of the city. In this respect, a theory of cities would both provide a language for understanding and interpreting urban data, and identifying parameters to act on to mitigate negative effects.

There are of course many studies that could fall in the scope of this Perspective, but I focus on specific examples that highlight the interdisciplinary aspects and the importance of physics tools and concepts, especially simple models with a physics flavour for which finding dominant mechanisms and critical parameters is the primary goal. The Perspective first describes datasets for testing models empirically, and then models for the size of cities, their spatial organization, human mobility and scaling hypotheses.

## II. Urban data

A way to understand complex systems familiar to physicists consists of constructing a microscopic model, leading to predictions at a larger scale that can be tested on empirical data. This feedback loop has led to many successes and relies crucially on the existence of empirical observations. The possibility of a quantitative analysis of social aggregate states is opened by the availability of large amounts of data about techno-social systems [8] (see for example the review about statistical physics of social dynamics [9]). In particular, information and communication technology has become a crucial source of data about cities, and these new technologies will also very likely affect the urban dynamics itself [10].

Datasets about cities can be classified in broad categories according to their time scale (Table 1 and Ref. [13]). At the scale of minutes to days, mobility data gathered by mobile phones, GPS, or RFIDs provide information about where and when people move in the city (for various uses and analyses of mobile phone datasets, see the survey [11]). This information reveals in depth the structure of activities in a city and patterns of mobility [12]. On longer timescales, from months to years, socio-economic surveys and censuses provide information such as total income, annual gasoline consumption and distance driven, and the relation between density and area, and can help understand and model various socio-economic processes. Finally, at the scales of decades and centuries, remote sensing and the recent digitalization of historical maps make it possible to study the evolution of urbanized areas, transportation and road networks, and to characterize quantitatively the importance of self-organization and planning in the evolution of urban systems. However, many data sources unfortunately have severe drawbacks: for example, mobile phone and GPS data are usually provided by private companies and are not publicly available and cannot be shared, and national surveys and censuses are of unequal availability and quality from one country to another, making comparative studies difficult — in addition to the lack of universal definitions of cities, described below.

## III. The size of cities

The population of cities varies over orders of magnitude, from small towns with hundreds of inhabitants to megacities with more than 10 million. Large cities are increasingly dominant and more than 40 megacities are expected to exist in 2020 [1]. This large disparity of sizes has been known for some time and in the 1940s George Kingsley Zipf uncovered the 'universal' behaviour of the form [13]

$$P(r) \sim r^{-\nu} \qquad (1)$$

where $P(r)$ is the population of the city at rank $r$ (cities are sorted by population in decreasing order). This Zipf's law is robust and valid for different periods in time: even if there is a non-trivial microdynamics with the rank of cities changing all the time, Zipf's law remains stable, in the sense that the exponent remains constant over large period of times (for the US, UK and the world, see [14]). The exponent $\nu \approx 1$ for most countries, but more recent empirical evidence suggests that there are non-negligible fluctuations of this exponent from a country to another (see Ref. [15] for statistical tests, Ref. [16] for an extensive study over 73 countries, and also Ref. [17] where sampling is shown to affect strongly Zipf's law).

The distribution of city sizes triggered many studies in economics and physics. How Zipf's law arises can be understood by starting with the first approach that describes the population evolution, the Gibrat model [18], which states that

$$P_i(t+1) = \eta_i(t) P_i(t) \qquad (2)$$

where $P_i(t)$ is the population of city $i$ at time (usually year) $t$. This equation describes the randomness of births and deaths by introducing an effective random growth rate $\eta_i(t)$ which is assumed to be independent between cities and uncorrelated in time. This simple equation with multiplicative noise naturally leads to a lognormal population distribution, in contrast with empirical results and Zipf's law. This simple model is therefore unable to explain Zipf's law and several other approaches have then been proposed, incorporating the effect of interactions between individuals [19] or variants of the Gibrat model. In particular, an approach that attracted much attention is based on the Gibrat model (equation 2) with the constraint that small cities cannot shrink to zero size [20]. In certain conditions this model leads to a power law with $\nu = 1$ (Ref. [21], [22]).

However, in reality cities can disappear and $\nu$ displays important fluctuations, suggesting that this model is not completely satisfying. The ideal approach would be based on reasonable mechanisms, linking individual decisions with the large-scale behaviour described by Zipf's law. An interesting step towards such a description — proposed in the context of the wealth distribution [23] — is to consider the diffusion equation with noise. This model has implications far beyond cities in phenomena such as finance, directed polymers and the KPZ equation. The diffusion equation with noise for the population $P_i(t)$ of city $i$, in the continuous-time limit, is

$$\frac{dP_i}{dt} = (\eta_i(t) - 1)P_i(t) + \sum_j J_{ij} P_j(t) - J_{ji} P_i(t) \qquad (3)$$

where the first term represents the 'internal' growth as given by Gibrat's law, the second term represents migration into a city, and the final term represents migration out of the city. The random variables $\eta_i(t)$ are assumed to be identically independent Gaussian variables with the same mean and variance given by $2\sigma^2$. The flow (per unit time) from city $i$ to $j$ is denoted by $J_{ji}$. For a general form of the $J_{ij}$ an exact solution of equation 3 is unknown. In the mean-field limit in which all cities are exchanging individuals with each other ($J_{ij} = J/N$, $N$ the number of cities), the stationary distribution $\rho$ of the normalized population $w_i = P_i/\overline{P}$ ($\overline{P}$ the average population) is [23]

$$\rho(w) \propto \frac{e^{-(\kappa-1)/w}}{w^{1+\kappa}} \qquad (4)$$

where the exponent is $\kappa = 1 + J/\sigma^2$. When the migration term $J$ is small (and non-zero), this regularization changes the lognormal distribution to a power law for large $w$ and for $J/\sigma^2 < 1$ the exponent is between 1 and 2. In this model, the exponent $\kappa$ is not universal and depends on the details of the system, providing a possible explanation for the diversity of values observed empirically [16]. This diffusion model suggests an interesting connection between a central model in statistical physics and the old problem of the urban population distribution. It also suggests that the origin of Zipf's law lies in the interplay between internal random growth and exchanges between different cities. An important consequence of this result is that increasing mobility increases $\kappa$ and therefore reduce the heterogeneity of the city size distribution. Other empirical tests are needed, for example on inter-urban migration datasets. From a theoretical point of view, it would be important to go beyond the mean-field analysis and to consider non constant couplings $J_{ij}$, a central problem in studies of the directed polymer model and the KPZ equation.

## IV. Spatial organization of cities

The diversity of available data makes it possible to study many problems about cities and their social and spatial organization. Even the quality of life and the vitality of a neighborhood can now be assessed thanks to new data sources [24, 25]. It is now possible to test qualitative ideas suggested by urbanists: for example in 1961, Jane Jacobs proposed four conditions to promote life in a city [26]: high density, small blocks, diversity of buildings and mixed land uses with different activity times. These guesses were confirmed for six Italian cities, using a combination of mobile phone data and socio-demographic information [24]. In the same spirit, the most 'vibrant' areas of London have been quantitatively defined and revealed using Twitter data as a proxy for human activity and smart-card data from public transport [25]. Social network datasets have a wide spectrum of applications including very practical ones, such as the prediction of the possible success of new venues for example [27].

### 1. Segregation and physics

Socio-economic considerations are the focus of many studies, both empirical and theoretical, and it is impossible to review all of them here. (See for example the review about statistical physics of social dynamics [9] and about models of collective behaviours [28].) Among these, social segregation — and in particular gentrification — is an important topic for cities, and it has been shown [29] that areas that have gentrified display similar morphological patterns. Modelling the dynamics of segregation is an important topic. Research on segregation dynamics started in the 1970s with the Schelling model [30], an Ising-type model — which naturally appealed to physicists. The model displays a rich phenomenology including static phase transitions, non-trivial coarsening, and complex behaviour when another type of agent is introduced [31] [32] [33] [34]. For more details see Box 1. In the Schelling model, agents are of two types and can move if they are not satisfied with their local environment. Within this model, even when individuals have a mild preference for their own kind, the system evolves towards a segregated state. The physics of this model was clarified in a numerical study [33] of a variant cast in terms of internal energy of particles [35].

In the Schelling model, and most of its variants, there is no collective optimization: each individual moves to improve its own utility, irrespective of the state of its neighbors. An individual move can therefore increase the number of 'unhappy' sites. That economical agents maximize their own utility contrasts with what usually happens in physics, in which systems of particles in equilibrium are in a state that optimizes a global quantity such as the energy. The link between microscopic ingredients and collective behaviour is elucidated by a Schelling-like model that interpolates continuously between individual and cooperative dynamics [31]. In this model, at each time step, an agent chosen at random moves to a vacant site with probability

$$P(\{\rho\} \to \{\rho'\}) = \frac{1}{1+e^{-G/T}} \quad (5)$$

where $\{\rho\}$ and $\{\rho'\}$ are the city-wide configurations of local densities $\rho$ before and after the move. The 'temperature' $T$ introduces noise in the decision process; if $T = 0$ the process is deterministic. The gain $G$ associated with the move is

$$G = \Delta u + \alpha(\Delta U - \Delta u) \quad (6)$$

and depends on both the variation $\Delta u$ of the agent's utility and the variation $\Delta U$ of the collective utility, both calculated for the specific move. The individual utility depends on the local density; the collective utility is the sum of individual utilities. The variations are coupled by $\alpha$, which measures the degree of cooperation between individuals. For $\alpha = 0$, $P$ depends on the agent's interest only (in the spirit of the Schelling model), whereas for $\alpha = 1$ the decision to move depends on the collective advantage only. The parameter $\alpha$ thus represents incentives created by a central government, for example. If the gain can be written as $G = V(\rho') - V(\rho)$, then detailed balance is satisfied and the system reaches an equilibrium described by a stationary probability distribution that can be computed. If such a function $V$ cannot be constructed, detailed balance is not satisfied and there are no general results known to date. However, this case is potentially very interesting to understand in a socio-economical context, and in particular the meaning of local equilibrium currents remains to be elucidated [28].

In the zero-temperature limit, for an overall city density equal to 1/2 and a particular choice of individual utility that is maximum for $\rho = 1/2$, there exists in principle a global social optimum for which all individuals are happy, because if $\rho = 1/2$ everywhere then all individual utilities are maximized [31]. Indeed, in the collective case ($\alpha=1$), the equilibrium corresponds to the maximum of the collective utility, given by $\rho_i = 1/2$ for all sites $i$ (Fig. 1a). In contrast, for the 'selfish' case ($\alpha = 0$), the equilibrium state is segregated (Fig. 1b): some of the blocks are empty and the others have a density larger than 1/2. This is obviously not an optimal situation but is a Nash equilibrium where no single move can improve the individual utilities (the stationary equilibrium corresponds to complete segregation $\rho = 0$ or $\rho = 1$; Ref. [28]). There is a transition at $\alpha = \alpha_c$ from a mixed to a segregated state as the cooperation decreases (or the influence of government incentives diminishes). The transition is driven by competition between the collective and individual components of the dynamics. The critical value $\alpha_c$ has been computed for particular choices of $u(\rho)$ [31]. This result is in sharp contrast with the 'invisible hand' theory of the 18th century economist Adam Smith which states that individual actions can result in unintended social benefits [28]. The model provides a counterexample in which individual optimization leads to a state far from the social optimum, even when the global incentives exist ($0 < \alpha < \alpha_c$).

## 2. Spatial distribution of activity

### 2.1. Datasets showing a polycentric transition

The spatial distribution of the economic activity in a city governs mobility patterns and the need for transport infrastructures and is thus a crucial factor in the organization of a city. The study of the spatial structure of cities has been prominent in urban economics [36] [37] and has led to many theoretical results, but often with weak empirical support. However, more recently, cell phone networks, which capture large amounts of human behavioural data [11], have provided information about the structure of cities [38] and their dynamical properties. For example, the activity density (that is, the density of individuals having a working activity) could be estimated from mobile phone data for 30 Spanish metropolitan areas [39, 40] and two types of cities were observed. One type is cities — typically small — with a unique activity centre. Such cities correspond to the classical image of the monocentric city organized around a central business district. The second type is larger cities with a more complex pattern of organization and more than one activity centre. Quantitatively, the number $H$ of activity centres ('hotspots') for a city with population $P$ is given by [40]

$$H \sim P^\sigma \qquad (7)$$

where the exponent $\sigma$ is around 0.5–0.6, a value confirmed by measurements of employment data for 9,000 US cities [41]. The number of hotspots thus scales sublinearly with population, a result that serves as a guide for constructing a theoretical model.

Earlier quantitative hints towards this polycentric structure were seen in a study showing anomalous scaling of the total traveled distance with the city population [42]. The polycentric structure was also observed with another mobile phone dataset [43], employment data [41], and was recently confirmed by using smart card transactions and taxi GPS trajectories [44]: using Singapore as a case study, taxi and public transport both reveal the polycentric structure of this city.

### 2.2. Modelling the polycentric transition

To explain why some cities are polycentric, I first discuss some classical models in urban economics [45] [46] for the structure of cities. A simple approach [45] describes the attractiveness of a given site in terms of a market potential. The time evolution of the density of companies is described by a nonlinear partial differential equation. Linear stability analysis shows that the uniform density is not stable, with the most unstable spatial mode $k^*$ depending on details of the system. For a 2D city of area $A$, the number of hotspots scales as $H \sim A k^{*2}$, which does not depend on the population (unless the area depends on the population). This simple model can explain the clustering effect of companies in cities but is not able at this point to explain the scaling of equation 7.

A more general framework, the Fujita–Ogawa model, takes into account the choice of residence and job locations for individual agents and companies, allowing in principle prediction of the spatial structure of renting cost and wages in cities [46]. More precisely, this model assumes that an individual chooses a residence located at $x$ and to work at location $y$ such that the quantity

$$Z(x,y) = W(y) - R(x) - T(x,y) \qquad (8)$$

is maximized. The quantity $W(y)$ is the typical wage earned at location $y$, $R(x)$ is the rental cost at $x$, and $T(x,y)$ is the transportation cost from $x$ to $y$, usually taken as proportional to the time $\tau(x,y)$ spent to cover this distance. A similar equation exists for the profit of companies. By neglecting congestion and taking the transportation cost as proportional to the Euclidean distance between $x$ and $y$, it can be shown that if the transportation cost for the typical interaction distance between companies is too large, the monocentric organization with a central business district surrounded by residential areas is unstable [46].

However, this general approach does not predict the resulting urban structure, but a simplified version yields testable predictions [41, 47]. In the simplified model, each agent has a residence located at random. Additionally, whereas in the real world wages result from a large number of interactions and factors, in the model they are replaced by a random number $\varphi(y)$, the distribution of which is not important: $W(y) = s\varphi(y)$ where $s$ sets the salary scale. This replacement is in the spirit of random Hamiltonians for heavy ions [48]. Finally, most journeys are made by car and the model includes congestion effects such that the travel time $\tau(x,y)$ depends on the traffic $Q(x,y)$ as described by the simple Bureau of Public Road function [49]

$$\tau(x,y) = \frac{d(x,y)}{\bar{v}}\left[1 + \left(\frac{Q(x,y)}{C}\right)^{\mu}\right] \qquad (9)$$

where $C$ is the road system capacity, and $\mu$ an exponent, usually between 2 and 5 (see, for example, Ref. [50]). These ingredients added to equation 8 enable a simple mean-field analysis showing that the monocentric organization is unstable for a value of the population larger than a threshold $P^*$, which depends on the details of the city. The number $H$ of distinct activity centers is given by [41]

$$H \sim \left(\frac{P}{P^*}\right)^{\frac{\mu}{\mu+1}} \qquad (10)$$

This simplified model thus predicts a sublinear behaviour for the number of activity centers with an exponent given by $\sigma = \mu/(\mu + 1)$. For $\mu \approx 2$, the empirical value $\sigma \approx 0.6$ is recovered. Whatever the value of $\mu$, the behaviour is sublinear, and congestion appears as a critical factor that shapes the structure of the city and favors the appearance of new activity centres.

Finally, knowledge of both residence and workplace locations allows discussion of multiple aspects of commuting to work [47]. In particular, the total commuting time, the quantity of $CO_2$ emitted by cars (discussed below), and the gasoline consumption [47] can be estimated.

## V. Mobility

Naively, the time spent on a given trip scales linearly with the distance traveled, but in fact the distance determines the mode of transportation — and therefore the velocity. Short trips are made using slow transportation modes with many stops, such as walking or public transportation. Longer trips are typically taken on fast trains or planes with a small number of stops. The relation between distance and travel time is therefore not simple [12]: it has been shown that the effective speed increases with the travel distance as a square root [51], owing to the hierarchical structure of transportation systems [52] (Fig. 2).

In the context of mobility within cities, it has been proposed that individuals have a travel time budget of about one hour per day, irrespective of the location or the epoch [53] [54] (see also the review [55] and references therein). When technology improves the speed, a larger distance can be covered within the time budget, allowing for the growth of cities and urban sprawl. In other words, individuals adjust their home and workplace in order to maintain an approximately constant travel time for their journey to work. This assumption was revisited empirically [56]: for some location the travel time can be stable while in others it can increase. Another study [47] confirms the increase of commuting time with population as a result of congestion, in sharp contrast with the travel time budget assumption. Apart from this general discussion, there are many studies on urban mobility, in particular using new data sources such as the choice of routes [57] [58] [59], universal patterns in human mobility [60], or congestion effects [61] [62]. The review [12] contains more references.

### 1. Mobility and energy

Moving an object requires kinetic energy and this physical consideration is expected to be relevant for human travel behaviour. The relationship between travel and energy was addressed in a study on a database for the UK spanning 26 years [63], which showed that there is a "constant energy expenditure" of the form $p_i \overline{t_i} \approx \overline{E}$ where $p_i$ is the energy consumption rate associated with transport mode $i$, $\overline{t_i}$ the average travel time for the transport mode $i$, and where the constant $\overline{E} \approx 615 kJ$ per person and per day. The same study also suggests the existence of a universal distribution of normalized daily travel time. An argument relying on entropy maximization with the constraint on the average energy being fixed at $\overline{E}$ leads to the canonical distribution for daily travel energies $P(E_i) \sim \exp(-\beta E_i)$, where for each individual $E_i$ is the daily energy spent using mode $i$ (Ref. [63]). However, the data [63] indicate that the probability of a small energy expenditure is vanishing, which can be accounted for by a cutoff of the form $\exp(-\alpha \overline{E}/E_i)$. This Simonson effect [63] implies that short trips are unlikely to be taken with a given mode because it is not worth to spend an energy amount less than $\alpha \overline{E}$ that can be interpreted as the energy necessary for the preparation of the trip. The energy distribution is then
$$P(E_i) \sim e^{-\alpha \overline{E}/E_i - \beta E_i} \qquad (11)$$
The existence of a universal distribution of travel times based only on physical variables such as times and energies highlights the importance of physical concepts for understanding mobility and social phenomena in general. In addition, the variables are directly measurable, in contrast with utilities introduced in classical choice modelling that describe preferences. However, this is a very simplified approach and many other ingredients should be integrated [68]. In particular, multimodal trips that combine different transportation modes are always more important in urban trips and are not described here.

### 2. Gravity law and radiation model

A first, apparently simple, question concerning mobility is to estimate the number of trips (per unit time) $T_{12}$ between two locations 1 and 2 of distance $d$ and of populations $P_1$ and $P_2$. An important model used in many applications and inspired from physics is the so-called gravity law in transportation research [64]

$$T_{12} = K \frac{P_1 P_2}{d^\tau} \quad (12)$$

Where $K$ is a positive constant and $\tau$ is a non-universal exponent characterizing these trips, and usually depends on factors such as the geographical area under consideration and the transportation mode. In the study reported in Ref. [64] $\tau = 1$. This formula was subsequently modified and there is now a large number of variants. Detailed discussions are found in a classic book [65]. An improved version of the gravity law was tested extensively against data [66] showing a good agreement. However, there are several theoretical drawbacks to the gravity law [67]. It does not have a clear derivation from microscopic mechanisms — there is a derivation based on entropy considerations but it fails to predict the dependence on $d$ (Ref. [68]). The law suffers from various problems such as divergence at small $d$, the absence of congestion effects and the absence of saturation when population increases, and it is deterministic, in contrast with fluctuations seen in empirical data.

The radiation model is simple, based on a small number of reasonable assumptions, aimed at overcoming these difficulties [67]. The model is named for its analogy with radiation processes in physics. In the radiation model, the focus is on the commute distance $r$: individuals are assumed to reside at a fixed location and then choose where to work. Individual expectations of job quality are characterized by a number encoding many factors such as income, working hours and conditions. This number is higher if the individual resides in an area with higher population, reflecting the fact that individuals in more populated areas have higher expectations. The surrounding locations offer jobs with a benefit that depends on the local population. In the same spirit as radiation and absorption processes, an individual living in $i$ will choose the closest job $j$ with benefit above their quality expectation, which leads to the average commuting flow from $i$ to $j$ (Ref. [67])

$$<T_{ij}> = T_i \frac{P_i P_j}{(P_i + s_{ij})(P_i + P_j + s_{ij})} \quad (13)$$

where $T_i$ is the total flow of individuals 'emitted' by location $i$, $P_i$ is the population in $i$, $P_j$ is the population in $j$ and $s_{ij}$ is the number of individuals in the disk centered on $i$ and of radius the distance $d(i,j)$. This central result shows that under some natural assumptions, the flow has the 'universal' form given in equation 13. It represents an alternative to the gravity law, has a clear derivation, relies on well-defined assumptions about the main mechanism (namely, how to choose a job), and is almost parameter free (this point has been discussed further elsewhere [69]).

The radiation model has been compared with data [67] [66] [70]. Improved versions of the gravity model can outperform the simple radiation model for small geographical units [70], but the main virtue of the radiation model is to provide a simple framework for discussing more complex mechanisms such as the relation between mobility and socio-economic factors. For example, one could ask if there is a relation between income and commuting distance $r$, and it seems that at least for three different countries (UK, US, and Denmark), the average commuting distance tends to increase with the level of income. It also displays a seemingly universal distribution slowly decreasing as $P(r) \sim r^{-\gamma}$ with $\gamma \approx 3$ [71]. This result appears as a consequence of an individual decision process close to the one described above [71]. A recent study [72] also explored empirically the income–commute relation and showed that richer individuals tend to take shorter trips in Singapore but longer in Boston. This might be due to the different organization of these cities: rich neighborhoods are peripheral in Boston and central in Singapore.

## VI. Scaling in cities

### 1. The hypothesis of scaling

The idea that a large city is a scaled up version of a smaller city is encoded in the hypothesis of scaling [73] [74]. This is a strong assumption that should be carefully scrutinized. In principle, it allows prediction of a property of a city from the knowledge of another city and the ratio of their populations. The important underlying assumption, which is intuitively reasonable, is that the population is a good determinant of cities, that is, cities with similar sizes share the same properties. The scaling hypothesis therefore leads to the idea that various aspects of cities can be explored by examining how a quantity $Y$ (which is usually extensive) varies with the population $P$ [73] [74]. The mathematical form for this scaling hypothesis is

$$Y \sim P^\beta \qquad (14)$$

where $\beta$ is an exponent, which in general is positive. The value of $\beta$ falls into three categories [75]. For quantities that depend on social interactions (such as the number of patents or number of serious crimes), $\beta > 1$. For example, empirical measurements [76] of the total number of mobile phone contacts and the total communication activity give an exponent around $1.1 - 1.2$. This superlinear scaling is possibly rooted in the fact that the number of interactions in cities grow rapidly with population, typically as $P^2$. In contrast, values $\beta < 1$ are observed when economy of scale is relevant, for example, for road surface or length of electric cables. A value $\beta = 1$ means that the quantity per capita $Y/P$ does not depend on the size of the city, which is the case for human water consumption or other human-dependent quantities.

Non-trivial values of $\beta$ suggest the existence of important mechanisms and could naturally serve as a guide for theoretical models. In this spirit, a general theory reveals that socio-economical quantities (such as wages or patents) increase superlinearly, with an exponent of the form $1 + \delta$ where $\delta$ depends on the fractal dimension of individual paths in the city; for most cases, $\delta = 1/6$ (Ref. [77]). These theoretical predictions agree with various empirical measurements [77], but not all of them [78] [79]. In addition, this theoretical approach is of a somewhat phenomenological nature, and it is difficult within its framework to connect microscopic ingredients and mechanisms to the collective emerging behavior. A model that makes it possible to understand and test the effect of various ingredients and mechanisms is still missing.

The hypothesis of scaling triggered a large amount of research activity that is ongoing, and also raises a number of questions [74] that are yet to be answered conclusively. (A recent criticism of scaling in cities was given in Ref. [80]).

### 2. The definition of cities

The first step in studying scaling is to determine a given quantity $Y$ and the population for a city. In general, the result depends on the city boundaries chosen. The quantity $Y$ is usually a sum over subunits in the city, and the scaling is thus relevant if the fluctuations in the subunits are not too large. Second, the exponent $\beta$ depends a priori on the city definition [74] [78]. The simplest choice is the administrative definition of cities which relies on well-defined boundaries. This is reasonable for smaller cities, but its meaning is less clear for larger urban areas which comprise in general a central core together with peripheral areas that are not captured by the administrative definition. Various other definitions of urban areas exist, but unfortunately, there is no consensus on them [74].

An unambiguous way to define cities is as the giant percolation cluster of the built-up area [81] (Box 2). A method that elaborates on this definition has been developed [78] and was used to construct for England and Wales a two-parameter family of cities with varying boundaries. For a given city, its central urban core is connected to surrounding areas that are sufficiently dense and large, and that exchange a number of commuters above a threshold. Varying the commuter and population density thresholds yields different city definitions. By comparing results for these definitions, the robustness of the exponent $\beta$ for various urban indicators was tested. This study shows that most urban indicators scale linearly with size, and when nonlinear scaling is present the value of the exponent fluctuates considerably (Fig. 3). For some quantities it can even be superlinear or sublinear depending on the city definition. This is a serious problem: the exact value of the exponent is not critical, but whether the behavior is superlinear or sublinear is crucial and should not depend strongly on the city definition.

The values of scaling exponents are also subject to fitting issues [79]. The data is usually noisy and incomplete and the number of available decades not very large. An important difficulty is that most values of $\beta$ are close to unity: typically for the superlinear quantities considered in Ref. [75] the exponent measured over two decades is in the range 1.07–1.34 with an average 1.19. The relevant question is then the existence of a nonlinear behavior [79]. When comparing a nonlinear fit of the form $aP^\beta$ with a linear fit $aP$, the nonlinear fit is always 'better', owing to its larger number of parameters (here, two compared to one), and a rigorous statistical analysis is needed to determine which fit is preferable. Deciding whether the best fit is linear also depends crucially on assumptions about the noise, which need to be tested. When this rigorous statistical analysis was performed over 15 different datasets, although the 'naïve' fit gave nonlinear behaviour the linear assumption could not be rejected in several cases [79]. This empirical problem is serious and should be addressed in future studies of scaling in cities.

## 3. Scaling and individual cities

Scaling for cities implies that it is possible — in principle — to predict the behavior of an individual city when its population changes. To discuss the relation between the scaling exponent and the individual city behavior, I focus on the particular case of delays due to traffic congestion with a dataset of 101 US cities in the time range 1982–2014 [82]. The scaling form obtained by aggregating the available data for different cities and for different years displays a nonlinear behavior, in qualitative agreement with empirical results [75]. However, the measured scaling exponent is unrelated to the dynamics of individual cities, which display a variety of behaviors and not a simple scaling law [82]. The delay seems not to be a function of population only, as is usually assumed for the scaling approach. The delay also displays aging, that is, it depends not only on the population but also on time, and possibly on the whole history of the city. This idea of path-dependency is natural for many complex systems such as spin glasses [83]. Indeed, it does not make sense in general to compare two cities of the same population but at very different dates: a population of one million means something different if it is in 1819 or in 2019 [82].

This discussion of congestion-induced delays shows that it can be problematic to mix data for different cities. Treating cities as scaled-up versions of each other is a strong assumption [84] and may be wrong for some quantities. Determining under which conditions scaling is correct is then a challenge for studies in this directions. Additionally, it has been recently suggested [85] that scaling exponents could experience time dependence and could converge to 'equilibrium' values, but more data are needed to test this assumption.

The abundance of urban data is beneficial for quantitative studies of cities, but this example shows that it is unclear if it is possible to use transversal data (that is, for different cities) to gain information about the temporal behaviour of individual cities. This is a fundamental problem that needs to be clarified when looking for generic properties of cities.

## VII. $CO_2$ emitted by cars in cities

A crucial question concerns the amount of $CO_2$ emitted by cities. Even if cities represent about 2% of the total land area, they release more than 70% of the total anthropogenic $CO_2$ emissions [86]. This emission is also concentrated in some urban areas: the 100 highest-emitting urban areas account for 18% of the global carbon footprint [87]. The determining factors of the carbon footprint are still under debate and there is not a simple relation with the population (Fig. 4).

Anthropogenic $CO_2$ has three major sources: transport, energy use in buildings, and manufacturing and industry. The relative proportion of each source depends on the population density, industrial activity and many other factors [88]. Most of the studies on this subject are empirical and often rely on purely statistical analysis, and a parsimonious model would be helpful for identifying the dominant mechanisms and critical factors.

Existing studies give conflicting results. A simplified variant of the Fujita-Ogawa model [47] shows that the $CO_2$ emitted by cars increases superlinearly with the population of the city, primarily because of congestion. In other words, larger cities are less green. This result stands in contrast with earlier work showing the opposite [89], in which the $CO_2$ emitted by cars was estimated using the distance traveled by commuters. However, the relevant quantity for car traffic is more likely the time and not the distance, and owing to congestion the two quantities are not proportional. Additionally, the choice of city boundary has a strong effect [90]: depending on its definition the scaling is sublinear or superlinear with population, leading to opposite conclusions about large cities and $CO_2$ emissions per capita. This uncertainty is reinforced by other studies reporting either linear [91] or superlinear [92] values of the exponent $\beta$ governing the scaling of the total $CO_2$. Furthermore, $\beta$ depends on additional factors and in particular decreases with the GDP per capita of the country a city belongs to [93]. In addition, a recent study of French cities suggested a parabolic behavior of $CO_2$ emissions versus population, implying the existence of a threshold above which a larger population leads to smaller emissions [94]. However, in that study, fluctuations were large. In general, car use and the corresponding emitted $CO_2$ is expected to depend on other factors such as the urban form and the possibility of using another transportation mode such as rapid mass transit. A recent study reports that the $CO_2$ emitted by cars is the product of three factors: the density of public transport, the typical size of the city and congestion effects [95]. In terms of scaling behaviour, the dominant term is proportional to population, and congestion induces a superlinear term but with a small prefactor. This combination might explain the difficulties in fitting this quantity with population.

This discussion shows the problem of urban $CO_2$ emissions is not yet well understood, and the other crucial problem of energy use in cities [96] could be discussed with the same conclusion. It is clear that new theoretical analysis are urgently needed to make sense of empirical data and to provide policy makers with robust arguments.

## Outlook

Many other aspects of cities have been studied quantitatively, such as the evolution of infrastructure networks [97, 98], the coupling between networks [99], multimodality [100] and heat islands and urban forms [101], in addition to the qualitative studies in social science. The examples discussed in this Perspective show how a combination of empirical results, economical ingredients and statistical physics tools can lead to parsimonious models with predictions in agreement with observations. In these examples, models for understanding the evolution of cities that agree with empirical observations can often be constructed. In particular, such models describe individual actions by stochastic processes and replace complex quantities resulting from the interactions of several agents with random variables.

However, the best approach for understanding complex systems such as cities remains an open question. Much debate centres on the place of theorization versus machine learning. Despite the undeniable successes of machine learning approaches for practical applications, these algorithms still act as black boxes and do not yet improve the fundamental understanding of the systems under study. High-level aggregation is needed in these methods and the approaches discussed here seem like viable alternatives for improving the theoretical understanding of complex systems. At a minimum, parsimonious models have the advantage of providing a simple language for making sense of the vast amount of data and identifying critical factors for the evolution of these systems. There is however a long way to go before being able to inform policy makers and urbanists with robust scientific arguments. I have argued here that data gathering and sharing, physical modeling and interdisciplinarity are important keys to achieve this crucial goal.

**Box 1: The Schelling model for segregation**

In the Schelling model, agents are described by Ising-like spins with tolerance level $f$ for other kinds. In 2D, when all agents are allowed to move without any spatial constraints as long as their situation does not get worse (see Ref. [33] for more details), the voids (empty sites) are expected to be randomly distributed in the system. If the vacancy density is too large, a cluster of void sites percolates through the system and prevents the growth of spin clusters. The clusters are then limited to the typical cluster size governed by the void density $\rho_0$. If instead $\rho_0$ is below the percolation threshold, the system converges to a quasi-ordered state with two large (macroscopic) domains spanning the whole system.

The dependence of the behavior of the system on the threshold $f$ is also interesting. For $\rho_0 \to 0$, segregation takes place even for large values such as $f=5/8$. For larger values $f \geq 6/8$, no segregation takes place and the system remains in a disordered state. Perhaps surprisingly, no coarsening takes place for extremely intolerant individuals: for $f=1/8$, the system remains trapped in a disordered state and cannot reach the optimal fully segregated state. This behavior is confirmed in the case of a variant of the Schelling model [32] where agents are satisfied if the number of unlike agents is lower than a fraction of all agents present in the neighborhood. Finally, a recent study [34] showed that a very small proportion of altruists — individuals with a high tolerance level — can modify the equilibrium state of this system.

**Box 2: Fractals, percolation and cities**

There is a long history between physics and the study of cities, and Zipf's law and the gravity law are important examples of this connection (see the main text). Other historical examples of such connections include models based on fractals and percolation. Fractals describe (disordered) structures that are self-similar on certain scales and were found in many different systems (see for example Ref. [102] and references therein). In particular, the fractal dimension of city boundaries was measured [103] [104] with a value between 1.2 and 1.4. The fractal dimension of the Paris subway network was also measured [105], and was found to be equal to 2 inside the city and to 0.5 in the suburbs. The diffusion limited aggregation model [106] was then invoked to describe the growth of such a fractal structures [103]. A later, alternative percolation model showed that correlations between newly added "buildings" predicts results in agreement with the measured dynamics of cities [107]. This correlated percolation model [108] is very simplified, but nevertheless suggested the possible relevance of simple statistical physics approaches for systems as complex as cities. The importance of percolation in cities was reinforced by a study that proposed a percolation-like algorithm to define cities without ambiguities as the giant percolation cluster of the built-area [81].

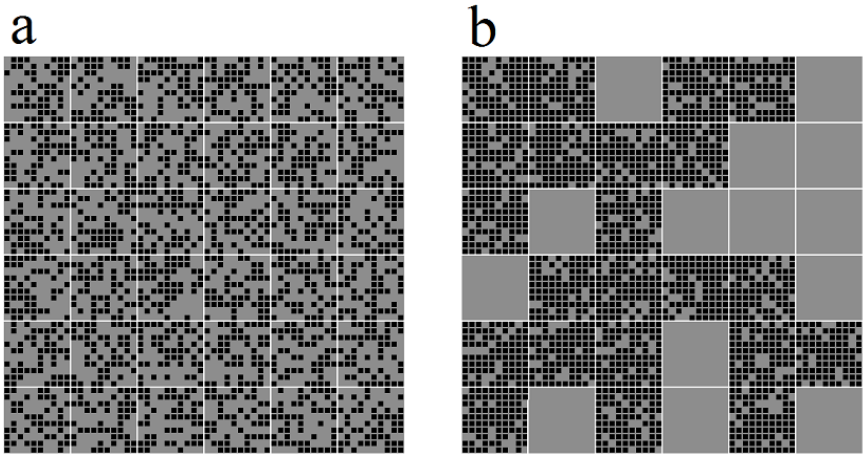

**Figure 1. Equilibrium configurations at zero temperature for cities with collective or selfish dynamics. a|** Collective optimization obtained when individuals move according to only the collective utility function. **b|** Segregation configuration obtained when individuals move according to only their individual utility function. In both cases, the overall city population density is 1/2 and the individual utility function is optimized for a local density of 1/2. Figure reproduced from Ref. [31].

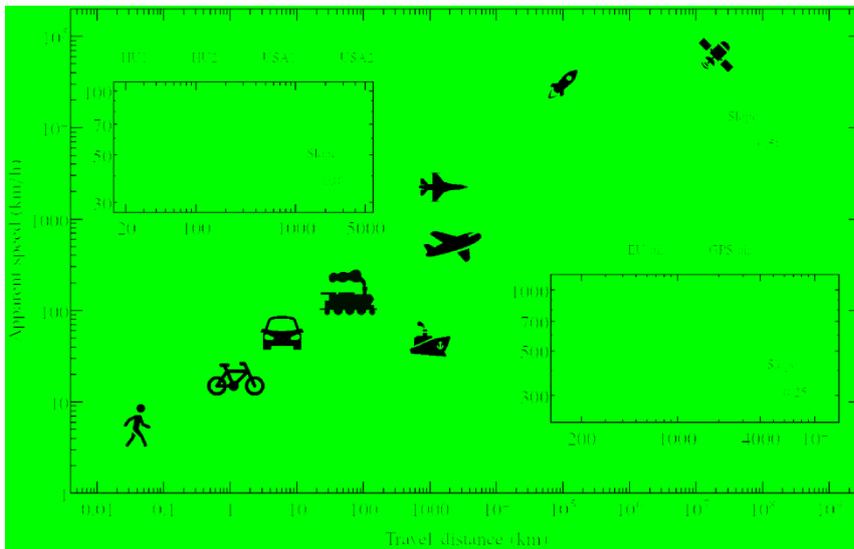

**Figure 2. Effective speed in human mobility as a function of travel distance.** Effective speed is computed as geodesic distance divided by travel time. Boxes signify the range of values measured for each mode of transport. The dashed line represents a square root behaviour (speed scaling with distance to a power of 1/2) as a guide to the eye. The insets present average results for car and air travel. For car travel, the scaling of speed with distance has an exponent 0.07; for air travel the scaling of speed with distance has an exponent 0.25 (as indicated by dashed lines in the insets). Figure reproduced from Ref. [51].

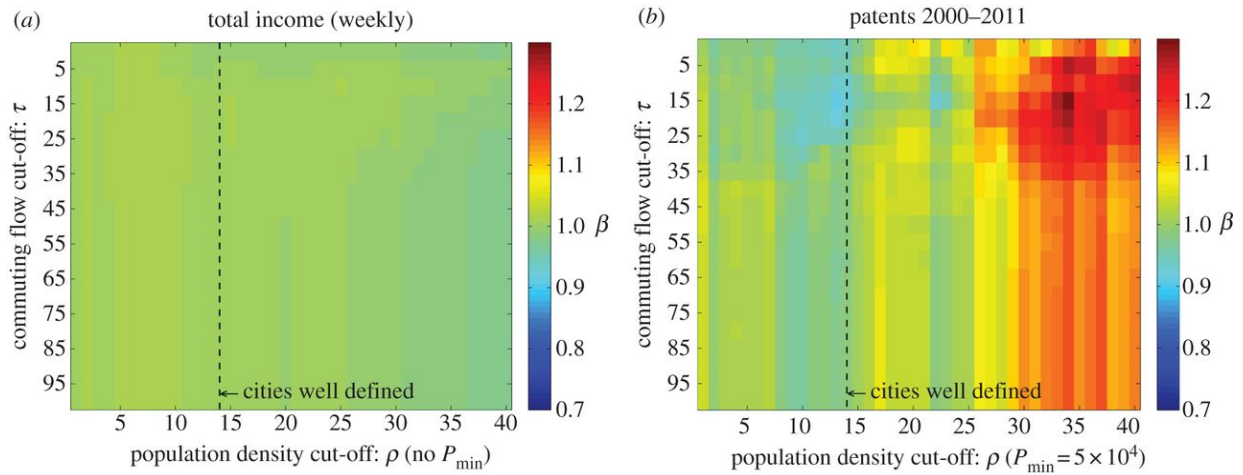

**Figure 3. Scaling of quantities with city population may depend on the definition of a city. a|** Heat map of scaling exponent $\beta$ for total income versus population. The behavior is consistently linear ($\beta = 1$) and does not depend on the city definition. No minimum population size is imposed, and cities are well-defined above a value of 14 inhabitants/ha for the density threshold. **b|** Heat map for the scaling exponent $\beta$ of the total number of patents for cities with more than $5\times10^4$ inhabitants. In this case the exponent can be smaller or larger than one depending on the city definition. In both panels, whether an area is part of a city depends on its population density being above a threshold $\rho$ and its commuting flow with the rest of the city being above a cutoff $\tau$. Figure reproduced from [78].

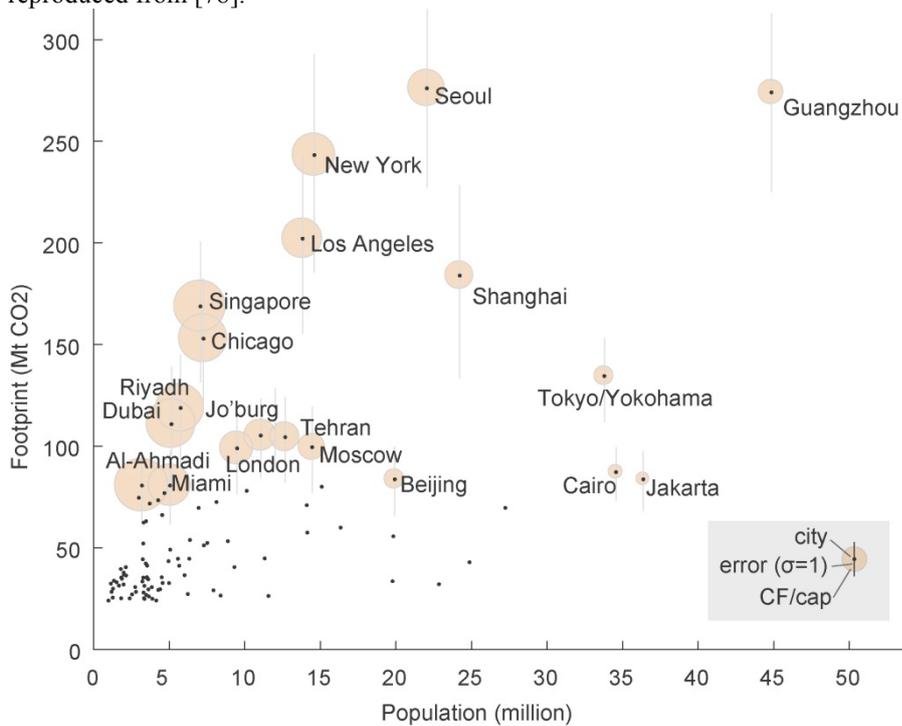

**Figure 4. Carbon footprint (in millions of tons of $CO_2$) versus population for various urban areas.** The named cities are the top 20. The size of the disk corresponds to the carbon footprint per capita and vertical lines to one standard deviation in the carbon footprint. Figure reproduced from [87].

**Table 1. Data sources according to their typical timescale and some phenomena occurring on these timescales.**

| Timescale | Data sources | Example phenomena |
|---|---|---|
| Minutes–days | <ul><li>GPS</li><li>Mobile phones (private companies)</li><li>RFID (transport companies such as RATP in Paris or TfL in London)</li></ul> | Spatial structure, mobility, urban activity |
| Months–years | <ul><li>Surveys</li><li>Censuses (US census bureau [109] or Eurostat [110])</li><li>City administrations</li></ul> | Social organization, housing market |
| Decades–centuries | <ul><li>Historical documents and maps (NYPL [11] or the geohistoricaldata research group [12])</li></ul> | Urban growth, self-organization, impact of planning |

Comments in parentheses are example data providers.